\documentclass{ifae}

\begin{document}

\begin{header}
  \title{Small $x$ QCD}

  \begin{Authlist}
    G.P.~Vacca\Iref{bo}

  \Affiliation{bo}
{Dip. di Fisica - Universit\`a di Bologna and INFN, Bologna, Italy}
  \end{Authlist}

  
  \begin{abstract}
A review of some theoretical aspects of small $x$ QCD physics is given,
with a particular emphasis to the relation between the BFKL and the colour
dipole approaches. The nonlinear evolution equations one may construct,
as a better approximation beyond the linear analysis, are discussed
together with their relation to a possible saturation regime.  
   \end{abstract} 
  
\end{header}

\section{Introduction}
Hadronic interactions at very high energies have been shown to be
driven by states with very high partonic densities. The first
phenomenological evidence for such states was obtained by the study of
DIS experiments at HERA where, for very small $x$, very high gluon
densities were found in the partonic description of the proton target.

The standard theoretical framework employed to study the parton densities and
structure functions of the proton in DIS is based on the QCD collinear
factorization, where the cross sections are decomposed in coefficient
functions (hard, perturbative) and parton densities (soft,
nonperturbative), the latter evolving according to the
DGLAP equations~\cite{DGLAP}. These linear equations
already provide the ingredients to a qualitative understanding of the
raise of the gluon density at small $x$, but the domain of
applicability is restricted by the assumptions adopted in their
derivation.

Essentially they are based on the requirement of large enough $Q^2$
values in order to neglect any kind of higher twist corrections and
the associated perturbative resummation is organized in powers
$\alpha_s^k (\alpha_s \log{Q^2})^n$, where the diagrams corresponding
to $k$ fixed and all $n$ generate the N$^k$L logarithmic
approximation.
At small $x$ one may take the DLA approximation of the DGLAP
evolution to pick up the dominant contribution to the gluon density
growth, corresponding to the contribution of the
$(\alpha_s \log{Q^2}\log{1/x})^n$ terms.

But to really extract the leading contribution at small $x$, if $Q^2$
is not extremely large, one may choose to resum all the
$(\alpha_s \log{1/x})^n$ contributions. This has to be associated to
the ``high energy factorization'' which decomposes the cross sections (and
amplitudes) into impact factors integrated with Green's functions in
the transverse space.
In particular the target impact factor integrated with the Green's
function corresponds to the unintegrated (in the transverse space)
gluon density. In this framework also a set of higher twist contributions
is taken into account.
This is known as the BFKL approach (LL)~\cite{BFKL},
which, as a matter of fact, predicts a very strong raise of the
gluon density, stronger than what is shown by the experimental
analysis in the $x$-window studied at HERA.
There are also two theoretical problems in the approach: the
appearance of an infrared instability related to a diffusion with the
rapidity evolution and the lack of running coupling description
withing the LL approximation. 

Successively an independent approach, equivalent to the LL BFKL,
named ``colour dipole picture'', has also been developed. It is useful
to understand certain aspects of small $x$ physics in a more economical way.

Further studies in the BFKL approach have been devoted to the NLL
case, where the resummation is extended to include contributions 
of order $\alpha_s(\alpha_s \log{1/x})^n$~\cite{NLLBFKL}.
The detailed analysis of these NLL contributions has shown that there
is a sort of instability due to I.R. singularities which do not agree
with the renormalization group flow. An estabished cure is to include
a set of corrections (of higher order with respect to the NLL scheme)
which are restoring the correct behavior. Within this context the
growth of the gluon density  and the cross sections $\sigma$ is characterized
by an intercept $\lambda\simeq 0.2$ ($\sigma \approx x^{-\lambda}$). 
The other piece of information needed in order to compute the cross
sections, the perturbative impact factors at NLL approximation,
are partially known. The Jet Verteces~\cite{BCV} have been derived,
while the virtual photon impact factor~\cite{BCGK} is in a advanced
status of computation. 

There is also available a montecarlo approach (CASCADE) where the small
$x$ regime is studied within the so called CCFM approach~\cite{CCFM},
which keeps ingredients of both the DGLAP and BFKL worlds in the LL
approximation. 

The main theoretical problem in common to the above described
approaches is that only a linear evolution is taken into account
with the resummation. 
This problem reflects the fact that unitarity is strongly
violated, in particular the Froissart bound.

In the perturbative approach the set of diagrams associated to
iterated effects of parton fusion or also of multiscattering (QM
shadowing) are completely neglected.

From the experimental point of view a new information has been
found, namely the observation that the DIS scattering data show,
within a very wide range of small $x$ and $Q^2$, a scaling
low~\cite{SGK}.
This regime is characterized by the appearance of a hard scale
(called saturation scale) and this fact is very interesting since it
may provide a perturbative scale in the high density regime at small $s$.
Moreover this scaling low, which exists also in the region where a
linear evolution is considered a reasonable approximation,
is implied if a saturation regime exists and provide a boundary
condition to the linear evolution equations.
 
It is in this direction that recently the small $x$ scientific
community has drawn his attention.
In particular the resummation of particular kind of parton fusion
effects has led to the formulation of non linear evolution
equations. Some ideas, previously pionereed and developed in the DLA
approximation~\cite{GLR}, have been developed in the colour dipole
picture (BK)~\cite{BK} and in the LL BFKL framework~\cite{num,NLBLV}.

The BK equation, which is the nonlinear equation for the colour dipole
evolution determined in the large $N_c$ leading approximation,
has been extensively studied both numerically~\cite{num} and, more
recently, analytically~\cite{analytic}, showing both the features of scaling
and saturation as a universal phenomenom.
The extension of this equation in the next-to-leading $1/N_c$
approximation has been given in~\cite{NLBLV} and one expects the same
qualitative behavior of the solution.

Still these approaches are only partially addressing the requirements of
unitarity and are therefore theoretically not very satisfactory, since,
beyond the fan structure which are typically generating the nonlinear
behavior, also some loop contributions are important.

The interest in the high density regime of QCD at small $x$ has been
raising in the last years also in connection with the heavy-ion
physics. A saturated regime, also called Color Glass
Condensate~\cite{CGC}, is seen as a possible prelude to a
thermalized phase, the Quark Gluon Plasma, which may exist in specific
cases for very short periods and is currently actively searched in the
experiments at RHIC.
 
\section{Linear Evolution: BFKL and Colour Dipole Physics}
Let us consider an high energy scattering process in the Regge limit
and within the perturbative regime ($s\gg -t \gg \Lambda_{QCD}$ in terms of
Mandelstam variables) in order to very shortly
illustrate some features of the high energy factorization and define a
link between the LL BFKL and the colour dipole approaches~\cite{NLBLV}.

The main result of the BFKL approach is that the leading
contribution to the cross section can be written as an integral in the
transverse space
\begin{equation}
\sigma \simeq \int d \mu_T \, \Phi_1 \, G(y) \, \Phi_2 \, ,\quad 
\frac{\partial}{\partial y}\,G=\delta +\frac{\bar{\alpha}_s}{2}K\, G
\, , \quad K=\omega_1 +\omega_2 +V \, ,
\label{HEF}
\end{equation}
where $y$ is the rapidity which plays the role of an evolution
parameter and $d\mu_T$ is the measure in the transverse space and $K$
is the BFKL kernel. The perturbative kernel $K$ is getting contributions from
virtual ($\omega_i$) and real (V is related to the square
of the Lipatov vertex for gluon production in the LL approximation)
with no I.R. singularities resulting from the sum.
The kernel may be seen as defining a quantum mechanical Hamiltonian
for two reggeized gluons, with the $\omega_i$ playing the role of the
kinetic term and $V$ the interation potential between them.
The Green's function $G$ defined above may be written when a
spectral basis of the operator $K$ is known. Usually in the literature
the BFKL equation is the eigenvalue equation for $K$.
The interacting two reggeized gluon system in a colour singlet state is
known as the perturbative BFKL Pomeron and the eigenvalues of the
kernel are related to its intercept. 
Let us note that the reggeized gluons are defined in a selfconsistent
way looking at the BFKL kernel properties in the colour octet state in the
$t$-channel.

The impact factors of colourless external particles possess the nice
property of being zero when the transverse momentum of one of the two exchanged
gluons is zero. This property is related to the gauge invariance and
let us have the freedom to look for different possible
representations for the space of functions of the impact factors and
for the domain of the operator $K$.
In particular one may consider the M\"obius representation when
restricts to a space of functions $f(\rho_1,\rho_2)$ such that
$f(\rho,\rho)=0$.
In such a case the BFKL equation is invariant under the M\"obius,
conformal, transformations.

The properties of the BFKL kernel in the M\"obius representations are
very interesting. The most important is the holomorphic separability,
which means that in the coordinate space, when defining a complex
$\rho=\rho_x + i \rho_y$, the kernel can be decomposed in a sum
$K=h+h^*$, where $h=h(\rho_i)$.

Starting from the Feynman diagrams derivation, in momentum space,
one obtains the following form for the BFKL kernel in terms of
pseudodifferential operators acting on functions with complex variables:
\begin{equation}
-K=H_{12}=
\ln \,\left| p_{1}\right| ^{2}+\ln \,\left| p_{2}\right| ^{2}+\frac{1%
}{p_{1}p_{2}^{\ast }}\ln \,\left| \rho _{12}\right| ^{2}\,p_{1}p_{2}^{\ast
}\,+\frac{1}{p_{1}^{\ast }p_{2}}\ln \,\left| \rho _{12}\right|
^{2}\,p_{1}^{\ast }p_{2}-4\Psi (1)\,.
\end{equation}
There are different possible ways to write this operator and the
alternative forms may differ depending on the space of functions
considered.
Among these forms there is one, for function related to the
M'\"obius representation, defined by a purely integral
operator, which concides with the kernel of the evolution of the
colour dipoles in the large $N_c$ limit~\cite{DP}:
\begin{equation}
K N=
\int \frac{d^{2}\rho _{3}}{\pi }\,\frac{\left| \rho _{12}\right| ^{2}}{
\left| \rho _{13}\right| ^{2}\left| \rho _{23}\right| ^{2}}\,\biggl(
N(\rho_{1},\rho_{3})+N(
\rho_{3},\rho_{2})
-N(\rho_{1},\rho_{2})\biggr) \,.
\end{equation}
The cross section in the colour dipole picture is written as
\begin{equation}
\sigma \simeq
 \int d^{2}\rho_{1}d^{2}\rho_{2}\,\int_{0}^{1}dx\,\left|
\psi(\rho_1, \rho_2;x)\right| ^{2}\,N(\rho_{1},\rho_{2}; y)\,,
\end{equation}
and the relation with the form in Eq. (\ref{HEF}) can be understood
on observing how to relate the impact factor $\Phi_1$ to the particle
wave functions, for example for a virtual photon~\cite{NLBLV}. One has
$\Phi_1=\int dx |\psi|^2 \theta_{IR}$, where $\theta_{IR}$ are the
phase factors which describes the four ways the two gluons attach to
the $q\bar{q}$ pair. The factor $\theta_{IR}\to 0$ when one of the two
gluons has zero 2-d transverse momentum. Nothing changes therefore
adding terms which are distributions having zero-support on that region.
Using such a gauge freedom one may write, in terms 
of an operator which projects onto the
M\"obius space of function, $\theta^{UV}$, such that
$\theta^{UV} f(\rho_1,\rho_2)=f(\rho_1,\rho_2)-1/2 f(\rho_1,\rho_1)
-1/2 f(\rho_2,\rho_2)$:
\begin{equation}
\sigma \simeq \Phi_1 \otimes G \otimes \phi_2=
\int dx |\psi|^2 \theta_{IR} \otimes G \otimes \Phi_2 =
\int dx |\psi|^2 \otimes \theta^{UV} G \otimes \Phi_2 = 
\int dx |\psi|^2 \otimes N \,.
\end{equation}
We have already reminded that the linear approches such as BFKL are
violating unitarity.
One may study in the LL approximation the linear evolution of more
complicated systems, with many reggeized gluons (more than 2) in the
$t$-channel as a first step towards the recovery of
unitarity~\cite{Bartels,KPJ}.

The homogeneous equation (BKP) for n-gluon colour singlet states are
governed by a kernel $K_n$ which is a sum of 2-gluon kernels.
Again the Green's function will be degined by
\begin{equation}
\frac{\partial}{\partial y}\,G_n=\delta +\frac{\bar{\alpha}_s}{2} K_n\, G_n\,.
\end{equation}
Even if the n-gluon impact factors have the property of vanishing
for a zero gluon momentum, the gauge freedom does not allow to
restrict all the solution to a space of function which are null when
two coordinates coincide. And, infact, there exist the leading
intercept Odderon solution~\cite{odderon} which does not posseses
this property.

Anyway there exist interesting states which belong to a (generalized)
M\"obius representation which are dynamically compatible with the BFKL
evolution and which will be considered in the following step towards
the unitarization, in section $4$.
But before proceeding in the illustration of some other interesting
phenomena appearing in the small $x$ QCD dynamics, we shall remind
some recent facts related to the strong interation, observed in the
DIS data measured at HERA. 
 
\section{Scaling in DIS Data}
Some interesting features have been recently found in the HERA DIS
data for the total cross section $\gamma^*p$~\cite{SGK}.
Infact it has been observed that
\begin{equation}
\sigma_{\gamma^*
  p}(Q^2,x)\approx\sigma\left(\frac{Q^2}{Q_s^2(x)}\right)\, .
\label{scaling}
\end{equation}
in the region $x\le 10^{-2}$ and $0.045 \le Q^2 \le 450 $.
The specific condition for this scaling to be realized is given by the
saturation scale depencence on $x$:
$Q_s^2\sim Q_0^2 \left( \frac{x_0}{x}\right)^\lambda$.

\begin{figure}[hbtp]
  \begin{center}
    \epsfig{file=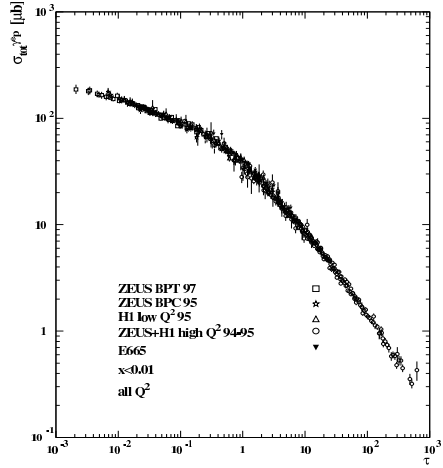,width=0.4\linewidth}
    \caption{The $\sigma_{\gamma^* p}$ HERA data in the kinematic range
$x\le 10^{-2}$ and $0.045 \le Q^2/$GeV$^2 \le 450$ are plotted with respect
to $\tau=\frac{Q^2}{Q_s^2(x)}$~\cite{SGK}.}
    \label{fig:ex1}
  \end{center}
\end{figure}

Therefore there exist a region where the general dependence on the two
kinematical variables reduces into a dependence on one new variable, resulting
from a combination of the two.
Analysing the $\sigma_{\gamma^* p}$ data with respect to the
$\tau=\frac{Q^2}{Q_s^2(x)}$ variable has indeed led to the plot of
Fig.~\ref{fig:ex1}.

A second interesting observation is that in the small $x$ region $Q_s$
constitutes an hard scale, since typically the parameters are such
that $Q_0=1$GeV for $x_0 \sim 3\times 10^{-4}$ and $\lambda \sim 0.3$.

This fact has been immediately considered associated to a possible
saturation in the gluon density for the kinematical region wherein
$Q^2 \le Q_s^2(x)$ but it has also been recognized that
the scaling behavior extends beyond the saturation region, up to
$Q^2 \ll Q_s^4(x)/\Lambda^2_{QCD}$~\cite{IIM}.
In this extra domain of scaling the linear evolution equations can be
used to describe the gluon density, and it has been shown that just
a saturated boundary condition has to be supplied to the linear
equations to recover the scaling behavior. 

\begin{figure}[hbtp]
  \begin{center}
    \epsfig{file=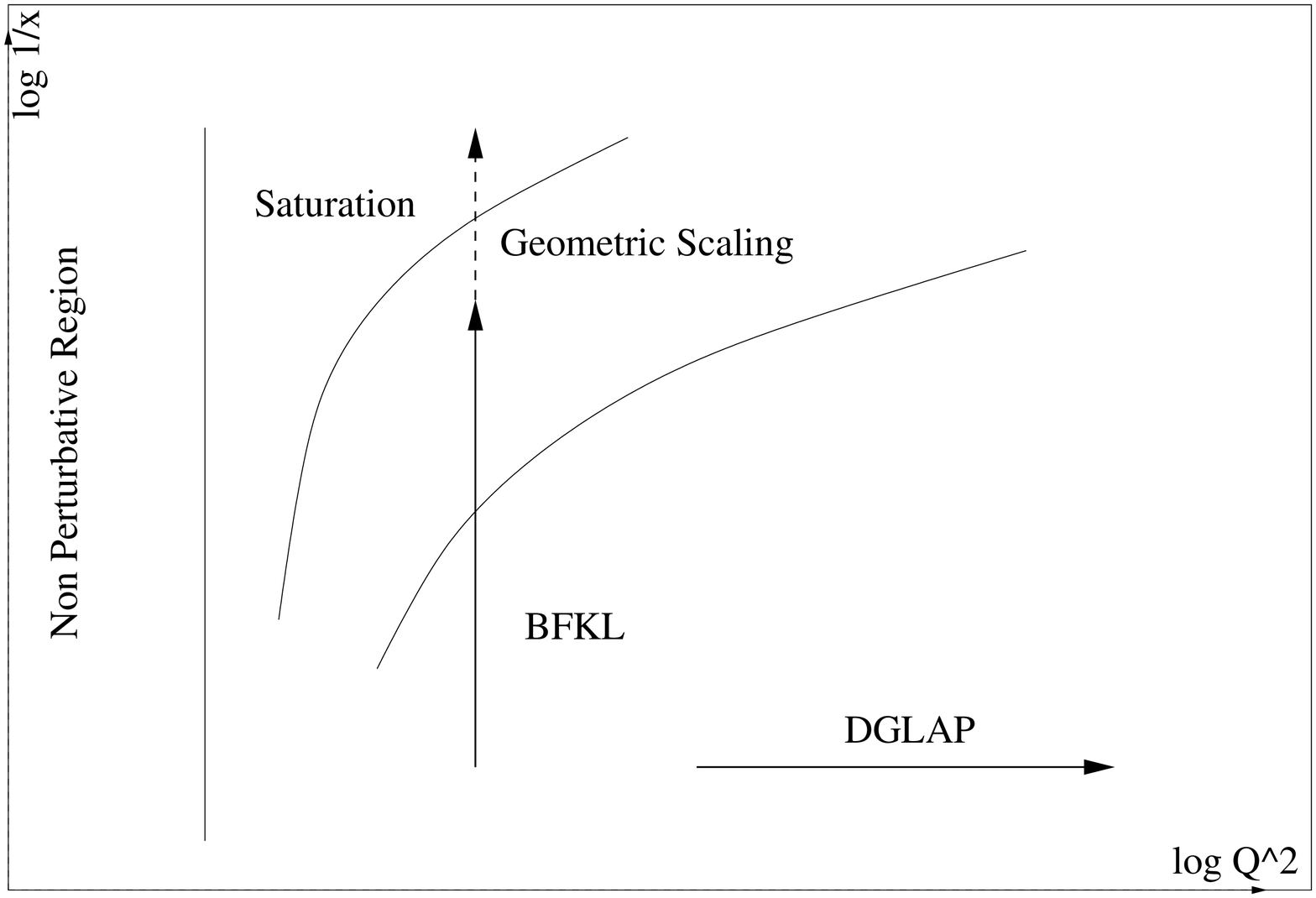,width=0.4\linewidth}
    \caption{Dynamical regimes in different kinematical regions of the
             $\log{Q^2}$-$\log{\frac{1}{x}}$ plane.}
    \label{fig:ex2}
  \end{center}
\end{figure}

The picture one may get for the different kinematical regions is
therefore described in Fig.~\ref{fig:ex2}, where the
$\log{Q^2}$-$\log{\frac{1}{x}}$ plane is presented.
\section{Non Linear Evolution: Leading Large $N_c$ and a Step Beyond}

Let us now consider the linear evolution of systems with different number
of gluons in the $t$-channel. Starting from the elementary verteces which
define the transition between states of different number of gluons in
the $t$-channel in the LL approximation~\cite{Bartels},
it is possible to organize a hierarchy of an infinite number of
coupled linear equations.

In particular these set of equations has been written and analyzed
explicitely in a systematic way for the systems of up to $6$
gluons~\cite{hierarchy4,hierarchy6}.  
In the $4$ gluon system one may extract an effective vertex
$V_{2\to4}$ which
defines the transition between $2$ and $4$ reggeized gluons. Any
solution can be decomposed~\cite{hierarchy4} in the sum of two terms,
$D_4^R$ and $D_4^I$, the latter being related to such a transition,
followed by an evolution governed by $G_4$.
The former term is governed by the gluon reggeization and different
choices can be made in the large $N_c$ limit~\cite{BV}.
In the large $N_c$ limit, when only colour planar diagrams dominates,
the effective vertex $V_{2\to4}$ becomes a simpler vertex which
decribes the splitting of a BFKL Pomeron into two. 

In the $6$ gluon system the solution of the coupled equations can be
decomposed in a sum of terms, one of them, $D_6^R$ is again related to
gluon reggeization, another contains a new effective vertex $V_{2\to6}$,
related to a transition where the colour structure of two Odderon is
present, and another may be seen as an iteration of two splitting of
the kind $V_{2\to4}$ in sequence in rapidity, each followed by the
BFKL evolution.

The diagrams with successive splitting in rapidity, denoted fan
diagrams, have been defined in the past in a different context~\cite{GLR}. 
Therefore one is tempted to consider all the diagrams, where
splitting in sequence are present, and to define an object which
describes a full resummation.
This approach has led to derive in a special case~\cite{num} the BK equation,
previously obtained by the resummation of colour dipole splitting in nuclear
targets~\cite{BK} in the limit $N_c\to \infty$. More recently, the fan
resummation has been reconsidered in order to investigate a general
relation between the BFKL and the dipole approaches and to define
an extension of the non linear evolution equation beyond the leading
large $N_c$ approximation~\cite{NLBLV}.

As usual, the non linear evolution appears when one is insisting in
defining an approximation of the full system, which is governed by linear
equations, in terms of a single smaller object and neglecting all kind
of higher correlations.

\begin{figure}[hbtp]
  \begin{center}
    \epsfig{file=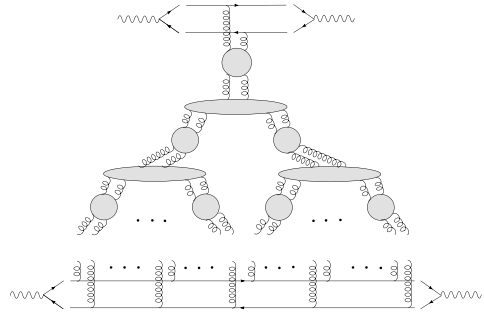,width=0.4\linewidth}
    \caption{Fan diagrams which are resummed, with the coupling of the
    gluons to the quark lines understood in all possible ways.}
    \label{Fig:ex3}
  \end{center}
\end{figure}

As a first step the leading fan structure is extracted on considering
the substitution $G_4 \to G_2 \otimes G_2$ so that one may write
\begin{equation}
\frac{\partial}{\partial y} \Psi=
\frac{\bar{\alpha}_s}{2} K \, \Psi - \bar{\alpha}_s^2 {\cal V}
\otimes \Psi \Psi \, ,
\label{eqfan}
\end{equation}
which resums the fan diagrams of Fig.~\ref{Fig:ex3}.
With ${\cal V}$ we have denoted the effective transition vertex.

Since the effective vertex correctly possesses the property, as the
impact factors, of being zero when any gluon line carries zero
momentum, one may perform a subtraction allowed by the gauge freedom and
so greatly simplify the vertex.
To this end we can rewrite Eq. (\ref{eqfan}) in the M\"obius space of
functions and choose the convenient rescaling
\begin{equation}
N_{\rho_1,rho_2}=8\pi \alpha_s \left(\Psi(\rho_1,\rho_2)-
\frac{1}{2}\Psi(\rho_1,\rho_1)-\frac{1}{2}\Psi(\rho_2,\rho_2) \right)
\, ,
\end{equation}
in order to obtain the BK equation
\begin{equation}
 \frac{d}{dy}N_{x,y}=\bar{\alpha}_s
\int \frac{d^{2}z}{2\pi }\,\frac{\left| x-y\right| ^{2}}{\left|
x-z\right| ^{2}\left| y-z\right| ^{2}}\,\left( N_{x,z}+
N_{y,z}-N_{x,y}-N_{x,z}N_{z,y}\right) \,.
\label{BKeq}
\end{equation}

This equation has been derived in the large $N_c$ limit approximation.
To go one step beyond and take into account the next-to-leading
$1/N_c$ corrections, one has to consider two kind of
contributions:
the one coming from the nonplanar term, subleading in $1/N_c$, in the effective
vertex and the one coming from the lowest $1/N_c$ correction to
the $4$-gluon Green's function, which at leading order is just the
product of $2$-gluon Green's functions, so that one introduces and
studies $N_4(\rho_1,\rho_2;\rho_3,\rho_4)=
N_{\rho_1,\rho_2}N_{\rho_3,\rho_4}+\Delta
N_4(\rho_1,\rho_2;\rho_3,\rho_4)$, which evolves according to
$\frac{\partial}{\partial y}\,H_4=\frac{\bar{\alpha}_s}{2} K_4\, N_4$.

Therefore it is possible to write a system of two coupled
equations~\cite{NLBLV}:
\begin{eqnarray}
\frac{d}{dy}N_{x,y}&=&
\bar{\alpha}_s
\int \frac{d^{2}z}{2\pi }\,\frac{\left| x-y\right| ^{2}}{\left|
x-z\right| ^{2}\left| y-z\right| ^{2}}\,
\Biggl[ N_{x,z}+N_{z,y}-N_{x,y}
-N_{x,z}N_{z,y} \nonumber \\
&&-\,{\Delta N_4(x,z;y,z)
- \frac{1}{2}\frac{1}{N_c^2-1}
\left( N_{x,z}+N_{z,y}-N_{x,y}\right)^2} \Biggr]
\nonumber \\
\frac{d}{dy} \Delta N_4 (\rho_1,\rho_2;\rho_3,\rho_4)
&=&
\frac{\bar{\alpha}_s}{2(N_c^2 -1)} \left(K_{12} + K_{34} \right)
\left(N_{\rho_1,\rho_3} N_{\rho_2,\rho_4} +
N_{\rho_1,\rho_4} N_{\rho_2, \rho_3} \right) \nonumber \\
&&+\, \frac{\bar{\alpha}_s}{2} \left(K_{12} + K_{34} \right)
\Delta N_4 (\rho_1,\rho_2;\rho_3,\rho_4)\,.
\end{eqnarray}

The next-to-leading $1/N_c$ corrections may provide a quantitative difference
from the leading description, but it is believed that the qualitative
picture will not change. Of great importance would also be to
investigate the NLL corrections (in $\log{1/x}$) to the above
equations.

The simpler BK equation (see Eq. (\ref{BKeq})) has been already analyzed
numerically showing the saturation phenomenon~\cite{num}.
For an analytical analysis it may be written, in a simbolic notation,
in momentum space as $\frac{\partial}{\partial y} \tilde{N}= \bar{\alpha}_s
\chi(\partial_{\ln k}) \tilde{N} - \bar{\alpha}_s \tilde{N}^2$ and
recently it has been studied in the quadratic approximation of the
complicated $\chi(\partial_{\ln k})$ pseudoifferential operator.

Infact in this way~\cite{analytic} it reduces to the Fisher and
Kolmogorov-Petrovsky-Piscounov (KPP) equation which, for reasonable
boundary condition (in agreement with colour transparency) admits
solutions which are travelling waves, i.e. functions which
depends only on a combination of the two kinematical variebles $Q^2$
and $x$ (or $y\sim \log(1/x)$), for large enough rapidities. 
They are characterized by a universal behavior described by
Eq. (\ref{scaling}) and by a general pattern of corrections.
 
\section{Conclusions}
Small $x$ QCD is showing, in certain kinematical regions, the
formation of a dynamical regime which can be partially studied with
perturbative field theory tools and was not expected a priory.
It is characterized by high parton densities and it has
been shown to lead to dynamical interactions which may be
described by non linear equations if one insists to keep simple
objects (gluon densities, perturbative Pomerons, or colour dipoles)
to describe the data.
The picture seems to be favoured by the DIS data which presents an
interesting scaling behavior in the small $x$ region.

As a general comment it is worth to note that most of the theoretical and
analyical work in the small $x$ QCD has to be seen as an effort to
better understand the QCD and the strong interactions.
A large fraction of the QCD community is tempted to use what is already known
as a tool in the challenge to understand the Standard Model (SM), the mass
generation mechanism (Higgs?) and the physics beyond the SM.
Nonetheless QCD still constitutes a challenge by itself since many dynamical
aspects still wait to be understood and therefore deserves further studies.


\begin{thebibliography}{9}

 \bibitem {DGLAP}
V.~N.~Gribov and L.~N.~Lipatov, Sov.\ Jour.\ Nucl.\ Phys.\ 
{\bf 15} (1972) 438;\\
  G.~Altarelli and G.~Parisi, Nucl.\ Phys.\ B {\bf 126}(1977) 298;\\
  Y.L.~Dokshitzer, JETP {\bf 46} (1977) 641.

 \bibitem{BFKL}
L.~N.~Lipatov, Sov.\ J.\ Nucl.\ Phys.\ {\bf 23} (1976) 338; \\
V.~S.~Fadin, E.~A.~Kuraev and L.~N.~Lipatov, Phys.\ Lett.\ B 
{\bf 60} (1975) 50;\\
I.~I.~Balitsky and L.~N.~Lipatov, Sov.\ J.\ Nucl.\ Phys.\ {\bf 28} (1978) 822;
\ JETP Lett.\ {\bf 30} (1979) 355.

 \bibitem{NLLBFKL}
V.~S.~Fadin and L.~N.~Lipatov,
Phys.\ Lett.\ B {\bf 429} (1998) 127; \\
M.~Ciafaloni and G.~Camici,
Phys.\ Lett.\ B {\bf 430} (1998) 349.

 \bibitem{BCV}
J.~Bartels, D.~Colferai and G.~P.~Vacca,
Eur.\ Phys.\ J.\ C {\bf 24} (2002) 83;
Eur.\ Phys.\ J.\ C {\bf 29} (2003) 235.
 \bibitem{BCGK}
V.~S.~Fadin and A.~D.~Martin,
Phys.\ Rev.\ D {\bf 60} (1999) 114008; \\
J.~Bartels, S.~Gieseke and C.~F.~Qiao,
Phys.\ Rev.\ D {\bf 63} (2001) 056014
[Erratum-ibid.\ D {\bf 65} (2002) 079902]; \\
J.~Bartels, S.~Gieseke and A.~Kyrieleis,
Phys.\ Rev.\ D {\bf 65} (2002) 014006; \\
J.~Bartels, D.~Colferai, S.~Gieseke and A.~Kyrieleis,
Phys.\ Rev.\ D {\bf 66} (2002) 094017.

 \bibitem{CCFM}
M.~Ciafaloni, Nucl.\ Phys.\ B {\bf 296} (1988) 49;\\
S.~Catani, F.~Fiorani, G.~Marchesini, Phys.\ Lett.\ B {\bf 234} (1990) 339;\\
S.~Catani, F.~Fiorani, G.~Marchesini, Nucl.\ Phys.\ B {\bf 336} (1990) 18;\\
G.~Marchesini, Nucl.\ Phys.\ B {\bf 445} (1995) 49.

 \bibitem{SGK}
A.~M.~Sta\'sto, K.~Golec-Biernat and J.~Kwieci\'nski,
Phys.\ Rev.\ Lett.\  {\bf 86} (2001) 596.

 \bibitem{GLR}
L.~V.~Gribov, E.~M.~Levin and M.~G.~Ryskin,
Phys.\ Rept.\  {\bf 100} (1983) 1.

 \bibitem{BK}
I.~I.~Balitsky, Nucl.\ Phys.\ B {\bf 463} (1996) 99; \\
Y.~V.~Kovchegov, Phys.\ Rev.\ D {\bf 60} (1999) 034008;
ibid. Phys.\ Rev.\ D {\bf 61} (2000) 074018. 

 \bibitem{num}
M.~Braun, Eur.\ Phys.\ J.\ C {\bf 16} (2000) 337.

 \bibitem{NLBLV}
J.~Bartels, L.~N.~Lipatov and G.~P.~Vacca, arXiv:hep-ph/0404110.

 \bibitem{analytic}
S.~Munier and R.~Peschanski, Phys.\ Rev.\ Lett.\  {\bf 91} (2003) 232001; \\
Phys.\ Rev.\ D {\bf 69} (2004) 034008.

\bibitem{IIM}
E.~Iancu, K.~Itakura and L.~McLerran, Nucl.\ Phys.\ A {\bf 708} (2002) 327.

\bibitem{CGC} See for example: \
E.~Iancu, A.~Leonidov and L.~McLerran,
``The colour glass condensate: An introduction,'', ``Cargese Lectures 2001'', 
arXiv:hep-ph/0202270 and references therein.

 \bibitem{DP}
A.~H.~Mueller and B.~Patel, Nucl.\ Phys.\ B {\bf 425} (1994) 471.

 \bibitem{Bartels}
J.~Bartels, Nucl.\ Phys.\ B {\bf 175} (1980) 365;\\
 \bibitem{KPJ}
J.~Kwiecinski and M.~Praszalowicz, Phys.\ Lett.\ B {\bf 94} (1980) 413;\\
T.~Jaroszewicz, Acta Phys.\ Polon.\ B {\bf 11} (1980) 965.

 \bibitem{odderon}
J.~Bartels, L.~N.~Lipatov and G.~P.~Vacca,
Phys.\ Lett.\ B {\bf 477} (2000) 178.

 \bibitem{hierarchy4}
J.~Bartels and M.~Wusthoff, Z.\ Phys.\ C {\bf 66} (1995) 157.

 \bibitem{hierarchy6}
J.~Bartels and C.~Ewerz, JHEP {\bf 9909} (1999) 026.

 \bibitem{BV}
M.~A.~Braun and G.~P.~Vacca, Eur.\ Phys.\ J.\ C {\bf 6} (1999) 147.
G.~P.~Vacca, PhD thesis, arXiv:hep-ph/9803283.

\end{thebibliography}
\end{document}